\documentclass{aa}
\usepackage{graphicx,epsf,psfig,epsfig,times,amsmath,amssymb}
\begin{document}
\title{First \textit{INTEGRAL} Observations of Eight Persistent
  Neutron Star Low Mass X-ray 
binaries\thanks{Based on
observations with \textit{INTEGRAL}, an ESA project
with instruments and science data centre funded by ESA member states
(especially the PI countries: Denmark, France, Germany, Italy,
Spain, and Switzerland), Czech Republic and Poland, and with the
participation of Russia and the USA}}
\author{A. Paizis\inst{1,}\inst{2},
V. Beckmann\inst{1,}\inst{3}, T.J.-L. Courvoisier\inst{1,}\inst{4},
O. Vilhu\inst{5},  A. Lutovinov\inst{6}, K. Ebisawa\inst{1},
D. Hannikainen\inst{5}, M. Chernyakova\inst{1}, J. A. Zurita Heras
\inst{1},
J. Rodriguez \inst{7,1},
A.A. Zdziarski\inst{8},  A. Bazzano\inst{9}, E. Kuulkers\inst{10}, T. Oosterbroek\inst{10}, F. Frontera\inst{11},
A. Gimenez\inst{12},  P. Goldoni\inst{13},
A. Santangelo\inst{14} and G.G.C. Palumbo\inst{15}}
\offprints{Ada.Paizis@obs.unige.ch}

\institute{\textit{INTEGRAL} Science Data Centre, Chemin d`Ecogia 16, 1290 Versoix, Switzerland
\and
CNR-IASF, Sezione di Milano, Via Bassini 15, 20133 Milano, Italy
\and
Institut f\"ur Astronomie und Astrophysik, Universit\"at T\"ubingen,
Sand 1, 72076 T\"ubingen, Germany
\and 
Observatoire de Gen\`eve, 51 chemin des Mailletes, 1290 Sauverny, Switzerland
\and
Observatory P.O. Box 14, Tahtitorninmaki, 00014 University of Helsinki, 
Finland
\and
Space Research Institute (IKI), High Energy Department, Ul. Profsojuznaya 
84/32,117810 Moscow, Russia
\and
CNRS, FRE 2591, CE Saclay DSM/DAPNIA/SAp, F-91191 Gif sur Yvette Cedex,
France
\and
N. Copernicus Astronomical Ctr., Bartycka 18, 00716 Warsaw, Poland
\and
CNR-IASF, Sezione di Roma, Via del Fosso del Cavaliere 100, 00133 Roma, Italy
\and
Research and Scientific Support Department of ESA, ESTEC, P.O. Box
299, NL--2200 AG Noordwijk, The Netherlands
\and
Dipartimento di Fisica, Universit\`{a} di Ferrara, Via Paradiso 12,
I--44100 Ferrara, Italy
\and
Instituto Nacional de Tecnica Aerospacial, Carretera de Ajalvir 4,
E--28850 Torrejon de Ardoz , Madrid, Spain
\and
CEA Saclay, DSM/DAPNIA/SAp, F-91191 Gif sur Yvette, France
\and
CNR-IASF, Sezione di Palermo, Via Ugo La Malfa 153, 90146 Palermo, Italy
\and
Universit\`{a} di Bologna, Via Ranzani 1, 40127 Bologna, Italy
}
\date{Received 14 July 2003 / Accepted 1 August 2003}
\authorrunning{Paizis et al.}
\titlerunning{\textit{INTEGRAL} First Observations of Eight Persistent
  Neutron Star Low Mass X-ray binaries}

\abstract{Early results from the INTEGRAL Core Program,
for a sample of eight persistently bright neutron star low mass X-ray binaries in
  the energy range from 5\,keV to  200\,keV are presented. 
It is shown that INTEGRAL efficiently detects sources and that spectra may be obtained 
up to several hundreds of keV by combining data from three of the four INTEGRAL 
instruments: JEM-X, IBIS and SPI.
For the source GX 17+2 it is shown that the spectrum extends well
  above 100 keV with a flattening of the spectrum above 30 keV. This might suggest 
a non-thermal comptonisation emission, but uncertainties in the current data reduction 
and background determination do not allow firm conclusions to be drawn.
\keywords{stars: neutron -- binaries: close -- X-rays: binaries --\textit{INTEGRAL} sources}}
\maketitle
%
\section{Introduction}


Since its launch in October 2002, the International Gamma-Ray Astrophysics Laboratory, \textit{INTEGRAL}, has been providing a large amount
of interesting data. The combination of the two wide
field of view (FoV) instruments, the imager IBIS (15\,keV -- 10\,MeV, $29^{\circ}\times 29^{\circ}$ partially coded FoV, Ubertini
et al. 2003) and the spectrometer SPI (20\,keV -- 8\,MeV, $35^{\circ}\times
35^{\circ}$ partially coded hexagonal FoV, Vedrenne et al. 2003) 
coaligned with the JEM--X
(Lund et al. 2003) and OMC (Mas-Hesse et al. 2003) monitors, allows large
areas of the sky to be observed and monitored in one pointing in a
wide frequency range from the optical to the $\gamma$-ray domain.
Such a capability is fully exploited during the \textit{INTEGRAL} Core
Program (a series of successive scans of the Galactic Plane (GPS; Winkler et al. 2003)
and Galactic Centre (GCDE)) which is regularly producing large amounts
of data, in particular on persistently bright sources.

The aim of this paper is to report preliminary results from early measurements
on eight persistent bright Low Mass X-ray binaries (LMXRB) hosting a
neutron star. The sample has been selected from a larger set  observed during 
the Core Program Observation scans on the Galactic Plane executed so far.
The  sources are listed in  Tab. \ref{table}. Given the type of
sources involved (about hundred of mCrabs in the 2--10\,keV band, Liu
et al. 2001) and the 
pointing exposures of about 2000 s, for single pointing spectral extraction, JEM--X 
for soft photons (5--20\,keV) and the low energy IBIS detector, ISGRI
(Lebrun et al. 2003) for harder photons (20--200\,keV) were
chosen. PICsIT, the hard photon IBIS detector (Di Cocco et al. 2003), has its 
peak sensitivity above 200 keV while the LMXRB are considered to display a
rather steep  spectrum and consequently have fluxes below PICsIT detectability.
The spectrometer SPI has been used to extract the hard energy spectra
(20-200\,keV) averaged on longer time scales. 

The combination of JEM-X, IBIS and SPI data provides a complete soft to hard
energy coverage, allowing a regular monitoring of source behaviour. Special 
attention is given to the hard emission ($>$ 50 keV) monitoring for
which \textit{INTEGRAL} will give unprecedented continuous coverage.

Sec. 2 of this paper gives an overview of the state of the art
studies on LMXRBs containing a weakly  magnetized neutron star and a
short description of the \textit{INTEGRAL} Core Program selected sample.
Sec. 3 contains details of the relevant \textit{INTEGRAL} observations
and data reduction methods used.
Preliminary results are summarised in the last Section.

\section{Bright LMXRBs in the \textit{INTEGRAL} Core Program}
\subsection{LMXRBs with weakly magnetised neutron stars}

LMXRBs hosting a weakly magnetised neutron star can be
broadly classified into two classes (van der Klis, 1995):
high luminosity/Z sources and Atoll sources covering a much wider
range in luminosity.
Z sources describe an approximate ``Z''-shape
(horizontal, normal, and flaring branch) in the colour-colour (CC) and
X-ray hardness intensity diagrams while Atoll
sources are characterised by an upwardly curved branch.
Two recent studies (Muno et al. 2002 and Gierli\'nski \& Done 2002)
suggest that the clear Z/Atoll distinction on the CC 
diagram is an artifact due to incomplete sampling: Atoll sources, if
observed long enough, \emph{do} exhibit a Z shape in the CC as
well. Many differences, however, remain: Atoll sources have weaker
magnetic fields (about $10^6$ to $10^7$\,G versus 10$^8$--10$^9$\,G of
Z sources),  are generally fainter ($0.01$--$0.3 L_{\text{Edd}}$
versus $\sim L_{\text{Edd}}$), can exhibit harder spectra, trace out the Z shape on longer time
scales than typical Z-sources and have a different correlated
timing behaviour along with the position on the Z.
Thus the distinction, at least from a practical point of view, still
makes sense.

Recent broad band studies, mainly with \textit{BeppoSAX}, have shown that many Z sources display a variable
hard power-law shaped component, dominating their spectra at $\gtrsim
30$\,keV (Di Salvo \& Stella \cite{disalvo1}; D'Amico et al. \cite{damico}; Di Salvo et al. \cite{disalvo2}, and references therein). This power-law has been explained as
non-thermal Comptonisation.
As previously said, lower luminosity systems (Atolls) can display much
harder spectra which can be well described with a simple power-law with photon indices 
of about 1.5--2.5.  Hard X-ray components
extending up to a few hundred\,keV have been seen in about 20 neutron
star LMXRB of the Atoll class. These sources usually display an exponential 
cut-off between many tens and a few hundred\,keV.  This component is interpreted as
 unsaturated thermal Comptonisation and is known as the ``hard
state'' of Atoll sources.

The long term X-ray variability of those sources has been extensively studied in the
2--12\,keV band with the \textit{RXTE} All Sky Monitor, as well as during \textit{RXTE}
dedicated pointings till about 40 keV (van der Klis \cite{klis3}; Swank \cite{swank}).
On the other side, \textit{BeppoSAX} pointings have shown the presence of the hard tails
mentioned above showing that Neutron Star systems, as well as Black Hole
ones, are capable of producing such hard photons.

The combination of regular monitoring in the hard X-rays
and $\gamma$-rays has not been done before and this is where
\textit{INTEGRAL} will give a major contribution to understanding the behaviour of bright Low Mass X-ray binaries from 5 keV up
to about 200 keV.

\subsection{The \textit{INTEGRAL} Core Program sample}
\noindent Thirtyfive per cent of \textit{INTEGRAL}  observing time, the Core 
Program, is time
reserved for the institutes that developed and delivered the
instruments, for the \textit{INTEGRAL} Science Data Centre (ISDC;
Courvoisier et al. 2003) and for the Russian scientists in return for 
providing the Proton rocket which put \textit{INTEGRAL} in orbit.

Fig. 1 shows the exposure map for a total of six months of
this program. As can be seen, the centre of our Galaxy has been
heavily covered, with the total observing time decreasing as we move away from it.
From the overall list of about 70 persistent and transient LMXRBs,
the 8 sources detected  primarily by IBIS/ISGRI and then by JEM-X
in at least 20 pointings, have been selected. At the time of writing 
OMC data on these sources is still quite sparse and, therefore, will not 
be included in this paper.
The red crosses in Fig. 1 show the position of those sources in the
Galaxy while in Table 1 the complete list is presented.
\begin{figure}
\vspace*{5mm}
\raisebox{2cm}{\includegraphics[clip=,angle=270,width=0.45\textwidth]{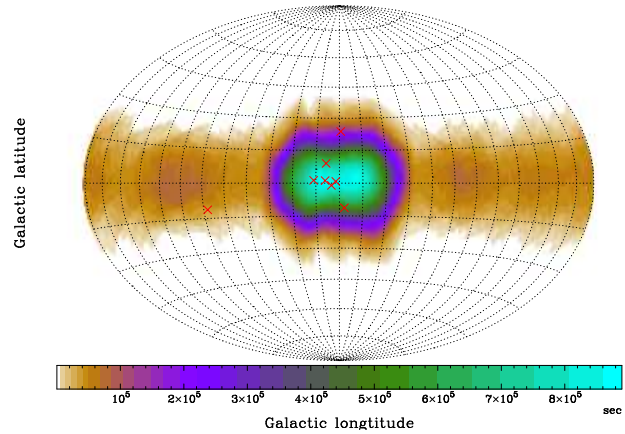}}
\caption{\small  Exposure map for a total of 6 months of GPS and GCDE data.  The
  spatial distribution of the 8  sources is visible (red crosses).}
\label{expo}
\end{figure}
\begin{table}[htbp]
\centering
\begin{tabular}{lrrccc}
Name      & \multicolumn{1}{c}{l} &
            \multicolumn{1}{c}{b} &
                                   Type &  $f_{\rm RXTE}$& $f_{\rm SPI}$ 
           \\
\hline
\hline
\object{Sco X-1}&359.09 & 23.78& Z &  11420 $\pm$ 1875 & 439 $\pm$ 4.1  \\
\object{1822-371}& 356.85 & -11.29 & ADC & 25 $\pm$ 29  & 33 $\pm$ 2.0 \\
GX 3$+$1&  2.29 &   0.79 &   AB & 390 $\pm$ 64 & 34 $\pm$1.9 \\
GX 9$+$9&  8.51 &   9.04 &  A   & 265 $\pm$ 44 & 31 $\pm$ 2.1  \\
GX 9$+$1&  9.08 &   1.15 &    A & 495 $\pm$ 75 & 34 $\pm$ 3.1 \\
GX 5$-$1&  5.08 &  -1.02 &    Z & 913 $\pm$ 137 & 77 $\pm$ 1.8 \\
GX 17$+$2& 16.43 &   1.28 &   ZB &  603 $\pm$ 104 & 56 $\pm$ 2.9 \\
\object{Cyg X-2}& 87.33 & -11.32 & ZB &  440 $\pm$ 80 & 35 $\pm$6.9 \\ \hline
\end{tabular}
\caption{Bright persistent LMXRBs regularly
monitored by \textit{INTEGRAL}.
  \emph{Type:} A=Atoll, B=bursting, Z=Z-source, ADC=Accretion Disc Corona;
   \emph{Flux:} average fluxes (in \text{mCrab}) as
  observed by
  RXTE/ASM (1.5 -- 12\,keV) and by  SPI (20 -- 40\,keV; about 1 Msec overall exposure). See text.}
\label{table}
\end{table}

 The exposure per source depends on its position relative to
the scan path and will differ from instrument to instrument due to
their different FoV. 

\section{Data reduction and analysis}
A large fraction of the Galactic Centre Deep Exposure (GCDE) has
already been completed.
One scan of the GPS is performed every 12 days on average.

We have analysed most of the  Core Program data currently available  with the
standard \textit{INTEGRAL} Data Analysis System (IDAS). 
The following 3 subsections provide the instrument specific analysis details. 
\subsection{SPI analysis}
The analysis of the SPI data is based on GCDE observations taken from
revolution 47 up to 62, i.e. between March 3rd and April 19th, 2003.
544 dithering pointings used in the analysis combine a total exposure
time of 958 ksec.
As the instrumental resolution of SPI
is about $2.5^{\circ}$, source confusion can affect the results. The ISGRI data have been used
as a reference to check for sources which might interfere in the SPI
data.
\object{GX 5-1} is separable by 40 arcmin from the black hole candidate GRS 1758-258. In this case
both sources will influence the results of source extraction of each
other and fluxes and spectra can only be taken as tentative. 
In the vicinity of \object{GX 17+2} two sources can be detected, 4U1812-12 and
AX J1820.5-1434, at $2.0^\circ$ and $1.2^\circ$ angular separation,
respectively. In these cases the SPI Iterative Removal Of
Sources (SPIROS) program (Skinner \& Connell, 2003) is able to disentangle the sources fairly well
as the closer one is rather faint, though minor effects cannot be excluded.   
SAX J1805.5-2031 is located $1.0^{\circ}$ away from \object{GX 9+1}. 

\begin{figure}
{\includegraphics[clip=,width=0.5\textwidth]{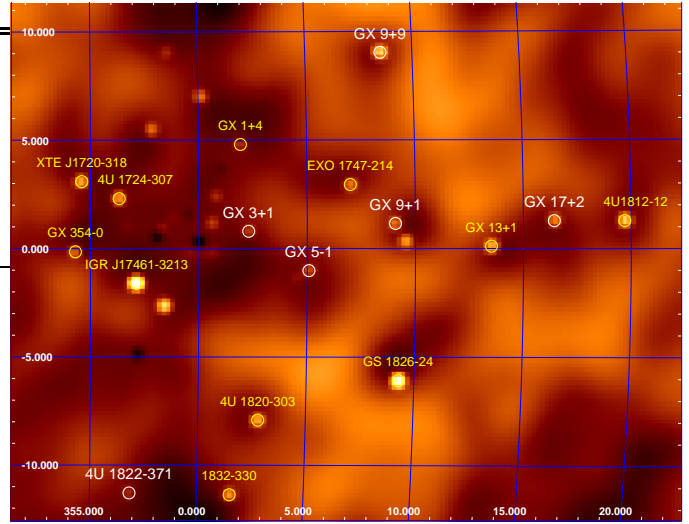}}
\caption{SPI image of the Galactic Centre in the 40 -- 100 keV band. The
  sources in white belong to our monitoring program. Other sources are
  marked in yellow for orientation.}
\label{spiima}
\end{figure}

As both sources appear to
have similar fluxes in the studied energy range, the spectrum of GX
9+1 derived from the SPI data might therefore be affected.
In the case of \object{GX 3+1}, two faint sources, SLX 1735-269 and {\mbox SAX
J1747.0-2858}, are at $2.2^{\circ}$ and $2.3^{\circ}$ angular separation,
respectively. As both sources are weak compared to GX 3+1, significant
effects on the spectral extraction of GX 3+1 are unlikely.
Fig. \ref{spiima} shows an image extracted from the SPI data in the
40--100\,keV region. In the very dense region of the Galactic
Centre, the extraction of sources seems to fail, while for more
isolated sources, like e.g. GS 1826-24 and GX 13+1, the results are
consistent with the ISGRI data (see Fig. \ref{isgriima}).  
For spectral extraction twenty logarithmic energy bins in the 20--200\,keV energy range have been applied to the data (Fig. \ref{spec}). The
instrumental response function used for the analysis, has been derived
from on-ground-calibration and corrected after the in orbit Crab
calibration observation. Source fluxes in the 20--40\,keV band
have been computed by comparison with results from Crab observation and
are therefore given directly in Crab units (see
Tab. \ref{table}). Fluxes in the 1.5--12\,keV energy band
have been extracted from the RXTE/ASM data base. The fluxes have been
averaged over the same time period as the SPI data in order to have
comparable results. In the case of 4U 1822--371 only the
measurements for which 
the flux added to the $1 \sigma$ error is larger than $0.0$ have been taken
into account, so that only reliable data are recognized. 
\subsection{ISGRI}
\begin{figure*}
\centerline{%
\begin{tabular}{cc}
\psfig{file=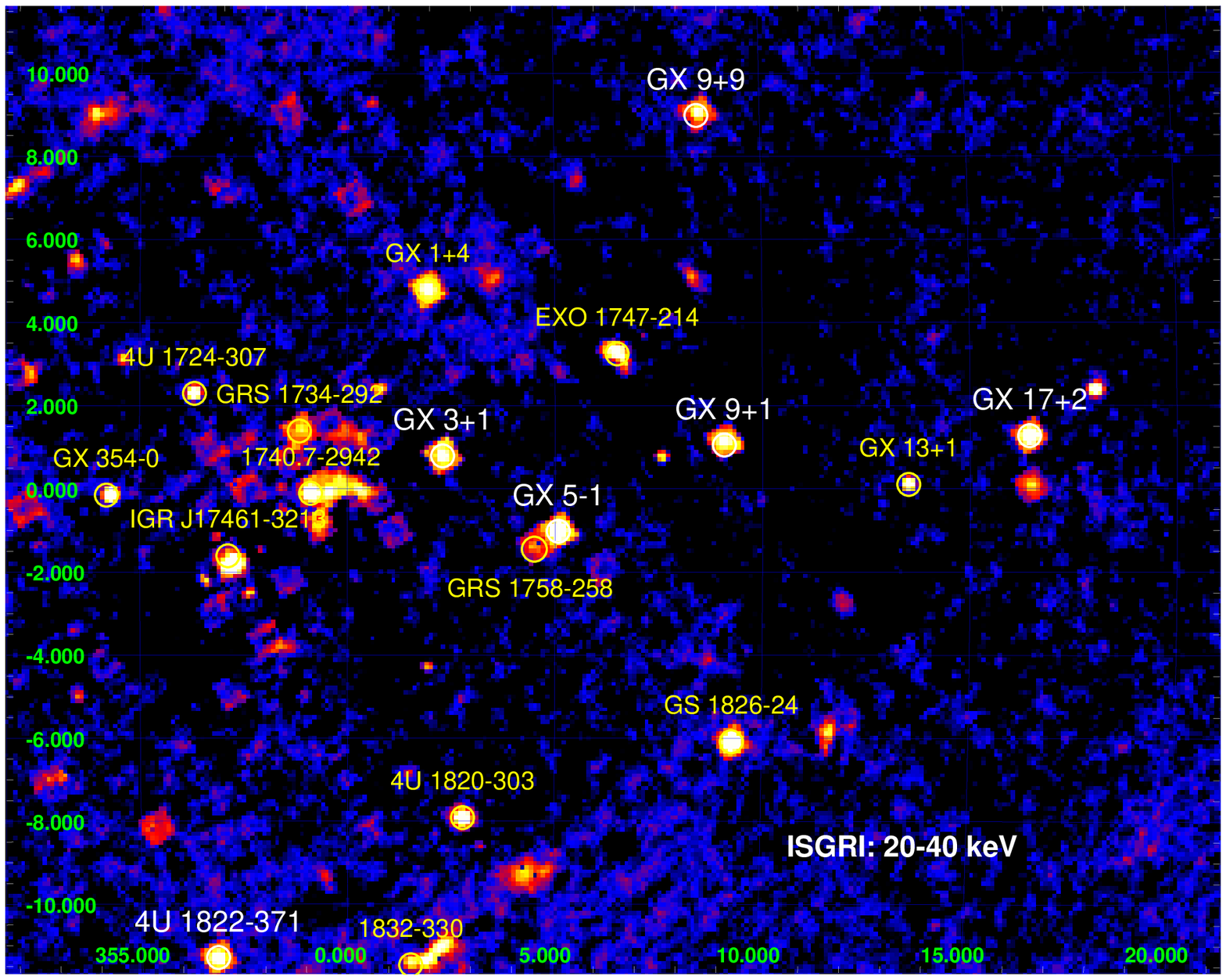, width=0.5\linewidth} &
\psfig{file=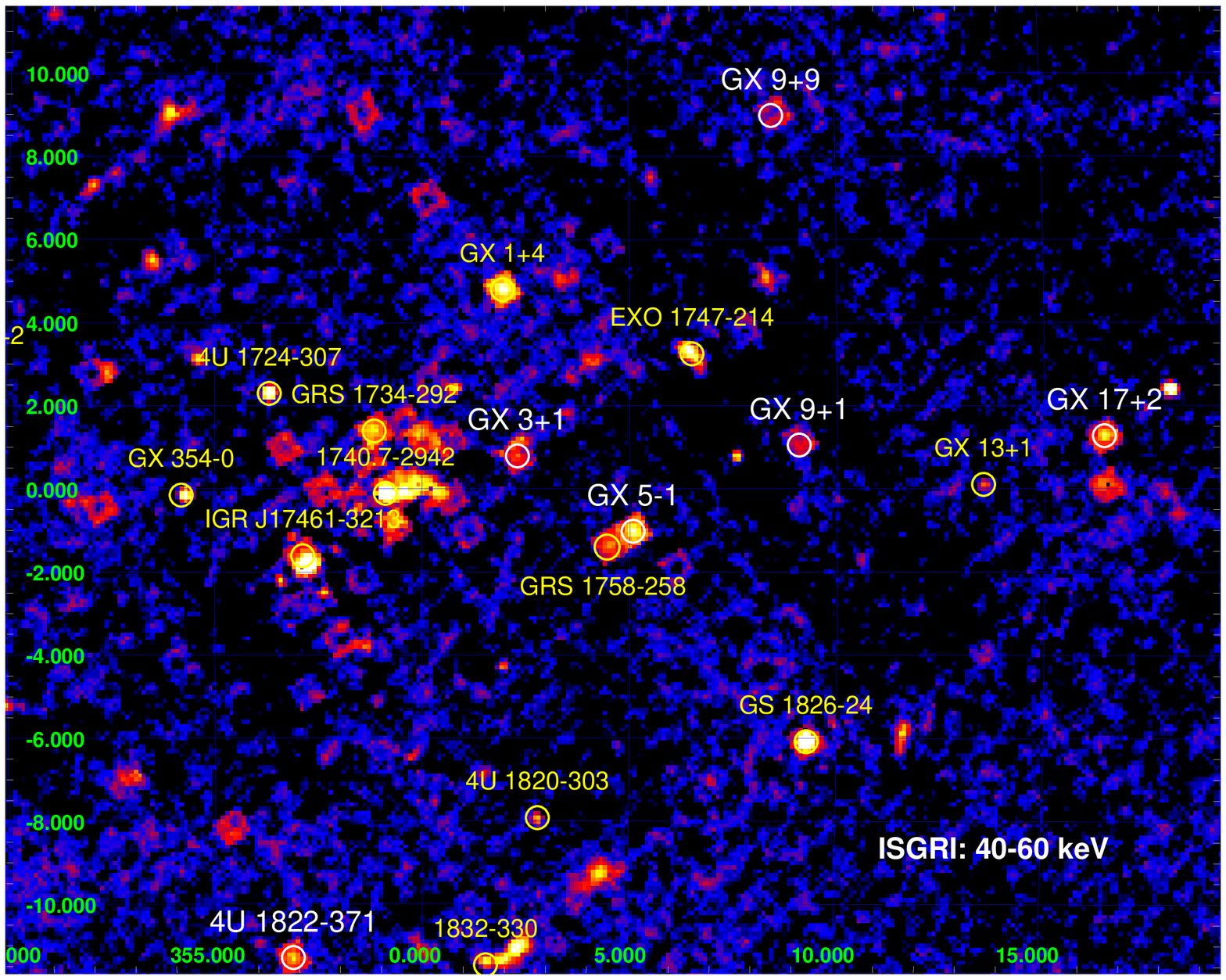, width=0.5\linewidth} 
\end{tabular}}
\caption{ISGRI images of the Galactic Centre in 20-40 keV and 40-60
  keV respectively. Not all the sources are labeled for clarity. The
  sources labeled in white belong to our monitoring program.}
\label{isgriima}
\end{figure*}


The analysis of ISGRI data is based on GCDE and GPS data from
revolution 30 to 64 i.e. January 11th to April 22nd, 2003.
One thousand pointings (for a total of about 2 Msec exposure\footnote{This is
  the total exposure time. The single source (point) exposure is much
  less as can be seen in Fig. \ref{expo}}) have been analysed separately and then
combined in the mosaicked image shown in Fig. \ref{isgriima}: a
zoom in the Center of the Galaxy in the 20--40\,keV and 40--60\,keV
bands.

\begin{figure}
{\includegraphics[clip=,angle=270,width=0.4\textwidth]{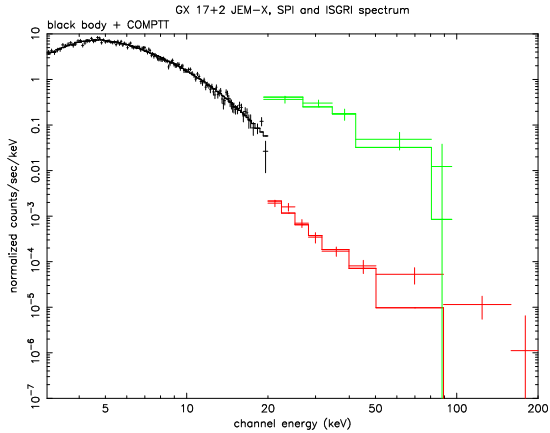}}
\caption{GX 17+2 combined JEM-X (black), ISGRI (green) and SPI (red)
  count spectra. The SPI spectrum is based on all GCDE data from March to
  April 2003, while
  JEM-X and ISGRI spectra are extracted from one pointing only (about
  2.2 ksec). Individual normalisation has been applied to the three
  instruments in order to compensate for uncertainties in the cross-calibration.}
\label{spec}
\end{figure}

Given the ISGRI sensitivity (5$\sigma$ detection in the 20--40\,keV band for a 20
mCrab source in one pointing of 2200 sec; Rodriguez et al. 2003, possible systematic errors are 
not taken into consideration) and  the
brightness of the sources of our sample,  we can extract
a source spectrum using (unlike for SPI) only data from one pointing. The
ISGRI spectrum for a single pointing
from revolution 54 for GX 17+2 is shown in Fig.~\ref{spec}.

\subsection{JEM-X}

The set of data taken from the same position for the same exposure time 
used for ISGRI has also been analysed for JEM-X using
the standard software. Only JEM-X2 data have been available during the
observations performed so far. The statistics for a bright LMXRB like GX 17+2
is of course even better than in  ISGRI (1 pointing from revolution 54
extraction) and the result is shown in Fig.~\ref{spec}.

\begin{figure}
{\includegraphics[clip=,width=0.4\textwidth]{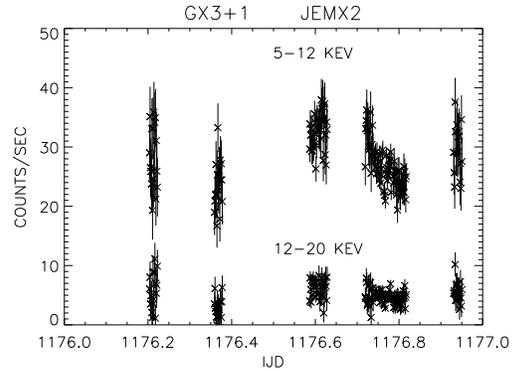}}
\caption{JEM-X light-curve for GX 3+1 in two energy bands. Each point
  refers to a 100 second time bin. IJD is the
  fractional number of days since January 1, 2000 (0 UT). See text for
  more details.}
\label{lc}
\end{figure}

Light-curves have also been extracted for each source and for each
pointing and a 1 day lightcurve for GX 3+1 is shown as an example in
Fig. \ref{lc} for two different energy bands (5--12\,keV and
12--20\,keV). For comparison the Crab pulsar shows a count rate in
JEM-X of $70 - 75 \, \rm{counts} \, \rm{s}^{-1}$ and   $20 - 25 \, \rm{counts}
\, \rm{s}^{-1}$ in the 5--12\,keV and
12--20\,keV energy band, respectively. Gaps in the data mean that for 
those periods the source was out of JEM-X FoV, the narrowest of
\textit{INTEGRAL} high energy instruments, thus there is no detection.

\section{Results}

Since its launch, \textit{INTEGRAL} has been producing a huge amount
of interesting data. Quite naturally at this early stage of the mission, however,
it is not possible to exploit them at best mainly because instrument
response functions are still under development and calibrations
on-going. More time is required before one can present long and
fast time scale variabilities in different energy bands as well as
detailed spectral behaviour of the sources. Nonetheless some main conclusions can
be already drawn. First of all, the combined spectrum in Fig. \ref{spec} shows the \textit{INTEGRAL} mission
wide energy band allows the emission of the binary system to be displayed all the way to the hard tail.
There seems to be no evidence for a cut-off below 100 keV. In this case the combined spectrum can be
represented by a simple model, including a blackbody plus a
comptonisation component.
The statistical quality of the fit does not justify adding an additional 
power law component. However, the present model flux is much below the 
measured one in the channel centered on 100 keV. This is a possible 
indication of the presence of a high energy tail on top of the blackbody 
Comptonised by thermal electrons. Physically, this can be realized if the 
electrons have a hybrid distribution, with a Maxwellian {\it and\/} a high 
energy tail. This appears to be the case in a number of black hole 
binaries (e.g., Gierli\'nski et al.\ 1999; Zdziarski et al.\ 2001).

Furthermore, based on one pointing only, JEM-X and ISGRI give 
a significant spectrum up to 100\,keV
for a source of $\sim 60 \, \rm{mCrab}$. This will allow the study
of the spectral time evolution on a pointing by pointing basis,
i.e. monitoring the spectral slope and cut-off on an hourly basis.
In this energy region the SPI sensitivity is much lower when compared
to ISGRI. The spectrograph can then be used to derive high energy spectra
based on longer observations. Though the spectrum is in this case an
average over the different stages of the LMXRB cycle, it can reveal
high energy tails, as seen in Fig. \ref{spec} up to $\sim 150 \, \rm{keV}$. This capability is also seen in the high energy (40 -- 100 keV)
image as shown in Fig. \ref{spiima}.

In addition, note that the ISGRI FoV is about the size of the image in
Fig. \ref{isgriima} so the amount of information one can obtain from one
ISGRI pointing is evident. Besides, its excellent angular resolution
(12' FWHM) allows the emission from close-by sources to be separated. See for
instance GX 5-1 and the 40 arcmin distant Black Hole Candidate GRS
1758-258, for which possible SPI source confusion has been pointed out. 

Another result of this study is the comparison of the SPI derived
fluxes to the RXTE/ASM fluxes as shown in Tab. \ref{table}. As both flux values have been extracted
over a long time period of 1.5 months, they represent the mean
status of the LMXRBs, i.e. averaged over all positions in the
colour-colour diagram. While the Z and Atoll sources of the sample show a steep
spectral slope between the 1.5--12\,keV and the 20--40\,keV band, the 0.59 second pulsar 4U 1822--371 (Jonker \& van der
Klis \cite{4U1822}) exhibits a comparably flat
spectrum with a photon index of $\Gamma \simeq 2$ between the RXTE and
SPI data. This hard spectrum has been reported before (e.g. Parmar et
al. \cite{4U1822hard}), but further investigations are necessary in
order to understand the nature of this illusive source.

Finally, though the X-ray monitor JEM-X covers a smaller area than the two
main instruments, Fig. \ref{lc} shows that monitoring of the X-ray flux on
short (several minutes) and long (days and months) time scale is possible.
This will allow to monitor outburst and the dipping behaviour of \object{4U 1822-371}.

\begin{acknowledgements}
We would like to thank Mike Revnivtsev for providing the mosaic tool
for the ISGRI images. We also thank P. Laurent and M. Cadolle-Bel for
kindly providing preliminary corrected ISGRI response matrices and the SPI colleagues 
for advice on how to handle SPI data. OV and DH acknowledge the Academy of Finland, TEKES, and the Finnish
space research program ANTARES for financial. J.R., acknowledges financial support from the French Space
Agency (CNES). A.A.Z., aknowledges KBN grants 5P03D00821, 2P03C00619p1,2, PBZ-054/P03/2001
and the Foundation for Polish Science. AP, AB, FF, AS, \& GGCP 
acknowledge ASI and CNR for financial support.

\end{acknowledgements}

\end{document}